\documentclass[prd,reprint,amsmath,showpacs,nofootinbib,superscriptaddress]{revtex4-1}

\usepackage[T1]{fontenc}
\usepackage[utf8]{inputenc}
\usepackage[brazil,english]{babel}
\usepackage{amsthm,amsmath,amssymb}
\usepackage{graphicx}
\usepackage[unicode]{hyperref}
\usepackage[usenames,dvipsnames]{color}

  \definecolor{darkblue}{RGB}{0,0,150}
\usepackage{graphicx}
\usepackage{tensor}
\usepackage{hyperref}
\hypersetup{
  pdfinfo = {
    Author = {Elcio Abdalla, Niayesh Afshordi, Michele Fontanini, Daniel C. Guariento, Eleftherios Papantonopoulos},
    Title = {Cosmological black holes from self-gravitating fields}
  },
  colorlinks = true,
  breaklinks = true,
  linkcolor = darkblue,
  urlcolor = darkblue,
  citecolor = darkblue
}
\usepackage{nicefrac}

\DeclareMathAlphabet{\mathbfit}{OML}{cmm}{b}{it}

\newcommand{\ud}{\ensuremath{\mathrm{d}}}
\newcommand{\linha}{\ensuremath{^{\prime}}}

\begin{document}

\title{Cosmological black holes from self-gravitating fields}

\author{Elcio Abdalla}
\email{eabdalla@fma.if.usp.br}

\affiliation{\foreignlanguage{brazil}{Instituto de Física, Universidade de São Paulo, Caixa Postal 66.318, 05315-970, São Paulo, SP}, Brazil}

\author{Niayesh Afshordi}
\email{nafshordi@pitp.ca}

\affiliation{Department of Physics and Astronomy, University of Waterloo, Waterloo, Ontario, N2L 3G1, Canada}

\affiliation{Perimeter Institute for Theoretical Physics, 31 Caroline Street North, Waterloo, Ontario, N2L 2Y5, Canada}

\author{Michele Fontanini}
\email{fmichele@fma.if.usp.br}

\affiliation{\foreignlanguage{brazil}{Instituto de Física, Universidade de São Paulo, Caixa Postal 66.318, 05315-970, São Paulo, SP}, Brazil}

\author{Daniel C. Guariento}
\email{carrasco@fma.if.usp.br}
\email{dguariento@pitp.ca}

\affiliation{\foreignlanguage{brazil}{Instituto de Física, Universidade de São Paulo, Caixa Postal 66.318, 05315-970, São Paulo, SP}, Brazil}

\affiliation{Perimeter Institute for Theoretical Physics, 31 Caroline Street North, Waterloo, Ontario, N2L 2Y5, Canada}

\author{Eleftherios Papantonopoulos}
\email{lpapa@central.ntua.gr}

\affiliation{Department of Physics, National Technical University of Athens, Zografou Campus GR 157 73, Athens, Greece}

\begin{abstract}

Both cosmological expansion and black holes are ubiquitous features of our observable Universe, yet exact solutions connecting the two have remained elusive. To this end, we study self-gravitating classical fields within dynamical spherically symmetric solutions that can describe black holes in an expanding universe. After attempting a perturbative approach of a known black-hole solution with scalar hair, we show by exact methods that the unique scalar field action with first-order derivatives that can source shear-free expansion around a black hole requires noncanonical kinetic terms. The resulting action is an incompressible limit of $k$-essence, otherwise known as the cuscuton theory, and the spacetime it describes is the McVittie metric. We further show that this solution is an exact solution to the vacuum Ho\v{r}ava-Lifshitz gravity with anisotropic Weyl symmetry.

\end{abstract}

\pacs{04.40.-b, 
04.20.Jb, 
04.70.-s, 
04.70.Bw
}

\maketitle

\section{Introduction}

Scalar fields, being the simplest tool in the hands of model-building physicists, are often used in the context of gravity and cosmology to build appropriate stress-energy tensors supporting models and to parametrize effects which could be more fundamental, but allow for a scalar description in certain regimes. Besides being easy to use, scalars, and in particular their appearance in fundamental theories, have recently become more ``natural'' following the discovery of the first observed scalar particle in nature, the Higgs boson.

Self-gravitating scalar fields play an important role in the formation of primordial black holes from inhomogeneities in the early Universe \cite{Khlopov:1985jw,*Rubin:2001yw,*Khlopov:2002yi,*Khlopov:2004sc}, and also as effective descriptions of fundamental fields in higher-dimensional black-hole spacetimes \cite{Maeda:2009ds,*Maeda:2010ja,*Maeda:2011sh,*Nozawa:2010zg,*Gibbons:2009dr}. Another important role played by scalars in the framework of black-hole physics has been to characterize the matter distribution outside the horizon, allowing for theoretical studies of the properties of black holes. There exist conjectures and theorems (the no-hair theorem in its many forms) forcing a scalar surrounding a black hole to take very specific configurations; for instance, a static black hole cannot have a nontrivial scalar hair on its horizon. On the other hand, hairy black-hole solutions were found \cite{bocharova-1970,*bekenstein-1974,*bekenstein-1975} in an asymptotically flat spacetime with the caveat that the scalar had to diverge on the horizon. Moreover, such solutions were later shown to be unstable \cite{bronnikov-1978}. 

In general, in order to  remedy the pathologies introduced by infinities, a regularization procedure has to be used to make the scalar field finite on the horizon. Roughly speaking, there are two ways to make the scalar field regular on the horizon, and they both  require the introduction of a new scale in the problem. The first possibility is to introduce a scale in the gravity sector of the theory through a cosmological constant; then hairy black-hole solutions were found in which all possible infinities of the scalar field were hidden behind the horizon \cite{Zloshchastiev:2004ny,Martinez:2002ru,*Martinez:2004nb}. The other way is to introduce a scale in the scalar sector of the theory. This can be done by adding in the Einstein-Hilbert Lagrangian a coupling between a scalar field and the Einstein tensor. This derivative coupling acts as an effective cosmological constant \cite{Sushkov:2009hk} and regular hairy black-hole solutions were found in asymptotically flat spacetime \cite{Kolyvaris:2011fk,Kolyvaris:2013zfa}.

Similarly, in cosmology scalar fields are also widely used. It would suffice to think of the inflationary paradigm, following which at early times a semiclassical scalar slowly rolls down its potential dumping energy in the gravitational sector of the theory, thus allowing spacetime to exponentially inflate. But the usefulness of scalars does not reduce to early-time physics; they are in fact the main ingredient of many proposals to explain late-time acceleration, among which we recall the quintessence models that employ a canonical scalar field coupled minimally to gravity with a scalar potential \cite{Fujii:1982ms,Ford:1987de,Wetterich:1987fm}, and the $k$-essence models in which the explanation of late-time acceleration relies on the nonlinear dynamics of a scalar field with noncanonical kinetic terms \cite{ArmendarizPicon:2000ah}.

Given that scalars seem to be ubiquitous in gravity and cosmology, it is interesting to investigate what the effect of a cosmological scalar field on a black-hole horizon is. In this case, the cosmological scalar field can vary with time over cosmological scales, and its behavior near the black-hole horizon cannot be inferred from the no-hair theorems which strictly apply only to stationary configurations.

Considering black holes surrounded by matter, the first example to consider is the well known time-dependent solution described by the Vaidya metric \cite{Vaidya:1951zz}. This is a spherically symmetric solution of the Einstein equations in the presence of a pure radiation field which propagates at the speed of light. The Vaidya solution depends on a mass function that characterizes the profile of radiation, and therefore any scalar field interacting with the Vaidya black hole should be massless \cite{Podolsky:2005ns}.

More realistic solutions can be found for spherical inhomogeneous dust, which are collectively known as the Lema\^{\i}tre-Tolman-Bondi (LTB) metric \cite{Lemaitre:1933gd,*Tolman:1934za,*bondi-1947}. However, the solutions generically fail upon shell crossing, which leads to formation of caustic singularities. Some aspects of black-hole dynamics in this context have also been studied \cite{firouzjaee-2010,Firouzjaee:2011hi}.

A more promising approach to the problem of finding black holes embedded in a cosmological background comes from looking for solutions representing black holes in a Friedmann-Lema\^{\i}tre-Robertson-Walker (FLRW) Universe. There have been many previous attempts to obtain such solutions, among which the Einstein-Straus model represents perhaps the simplest one \cite{Einstein:1945id}. In this model the universe is built by patching Schwarzschild black-hole spheres with a FLRW metric filling the rest of the volume. However, these black holes are time symmetric, and so they fail to describe an overall dynamical black-hole configuration in the Universe. Similarly, the ``Swiss cheese'' models patch FLRW pressureless spacetime with spherical holes containing compensated LTB solutions. These models have recently gained popularity as idealized toy models for cosmological observations in an inhomogeneous universe (see, e.g., Ref.~\cite{Marra:2007pm}).

Finally, one can consider the McVittie solution \cite{mcvittie-1933}, a spherically symmetric, time-dependent solution of the Einstein equations that describes a black hole embedded in a FLRW universe when the expansion asymptotes to de~Sitter. The solution approaches a FLRW metric at large distances, while it looks like a black hole at short distances (at a horizon scale), and, despite the criticisms raised in the literature that the McVittie spacetime contains singularities on the horizon \cite{Nolan:1998xs,*Nolan:1999kk}, recently it was shown \cite{Kaloper:2010ec} that if a positive cosmological constant is introduced, then the metric is regular everywhere on and outside the black-hole horizon. Similarly, the structure of the McVittie solution with a negative cosmological constant was discussed in Ref.~\cite{Landry:2012nv}.

The usual approach followed to study the effect of a cosmological scalar field on a black hole is to use a probe approximation in which the scalar field does not backreact onto the metric; i.e., the scalar is assumed to evolve in a static background \cite{Jacobson:1999vr,Frolov:2002va,Saravani:2013kva}. An improved probe ``slow-roll-like'' approximation was used in Ref.~\cite{Chadburn:2013mta} in which a time-dependent scalar field evolves in a time-dependent black-hole background.

In this work, however, we aim at addressing the fully self-consistent dynamical problem. In Sec.~\ref{sec:mtz-mcv}, we start by considering a perturbative approach to a black-hole solution generated by a scalar field conformally coupled to curvature; we show that for sufficiently smooth deviations from such a solution the system of coupled Einstein and scalar-field equations admits no solution. In Sec.~\ref{sec:cuscuton_all}, we then investigate compatibility of most general scalar field actions with first-order derivatives  (otherwise known as $k$-essence) with the McVittie spacetime, which singles out an incompressible limit of these theories, known as quadratic cuscuton \cite{Afshordi:2006ad,*Afshordi:2007yx}.  Finally,  Sec.~\ref{sec:conclusions} concludes the paper.

Throughout the paper, Greek indices run from 0 to 3 and we use the $(-,+,+,+)$ signature. We denote partial derivatives with respect to $t$ by a dot and with respect to $r$ by a prime.

\section{Cosmological black holes with standard fields} \label{sec:mtz-mcv}

Consider the  McVittie spacetime~\cite{mcvittie-1933} in isotropic coordinates:
\begin{equation}\label{mcvittie}
  \ud s^2 \! = -\!\left( \frac{1 - \frac{m}{2 a r}}{1 + \frac{m}{2 a r}} \right)^{\! 2} \!\! \ud t^2 \! +  a^2 \! \left( 1 \!+\! \frac{m}{2 a r} \right)^4 \!\! \left( \ud r^2 \!+\! r^2 \ud \Omega^2 \right) \! ,
\end{equation}
with $a = a (t)$ and $m$ constant. This metric, presented many decades ago, is a classical solution of the Einstein equations which may describe a black hole (or a black-hole/white-hole pair) in a shear-free cosmological background \cite{Kaloper:2010ec,Lake:2011ni,Faraoni:2007es,Carrera:2009ve,Faraoni:2012gz,daSilva:2012nh}. The classical source that generates this geometry is a comoving nonbarotropic perfect fluid with homogeneous energy density \emph{but inhomogeneous} pressure. It is the unique perfect-fluid solution of the Einstein equations which is spherically symmetric, shear-free, and asymptotically FLRW with a singularity at the center \cite{raychaudhuri}. In the limit $r \gg \frac{m}{2 a}$, the McVittie spacetime asymptotes to a flat FLRW universe with scale factor $a$. If $a$ is set to be constant (say, $a \equiv 1$), we recover the Schwarzschild spacetime with Arnowitt-Deser-Misner mass $m$ written in isotropic coordinates. After a lengthy discussion in the literature, it has been proved that, at least in some cases (that is, when the scale factor tends to that of de~Sitter space), the line element~\eqref{mcvittie} describes a black hole immersed in a FLRW universe \cite{Kaloper:2010ec,Lake:2011ni}.

A classical general relativity solution is of course interesting on its own, but it can lead to even more interesting results when it can be shown to be the gravitational counterpart of a field-theoretical model. This, as will be extensively presented later, is the case for the McVittie solution \eqref{mcvittie}.

Inspired by its mathematical simplicity, one might ask whether metric \eqref{mcvittie} or some generalization thereof is a solution of the system composed of a scalar field coupled to gravity. However, a minimally coupled comoving canonical scalar cannot be the answer, since it has homogeneous density and pressure, contrary to what is needed to source McVittie. Considering a richer action, say, by adding a vector field, also does not help due to another characteristic of metric \eqref{mcvittie}, namely, spatial Ricci isotropy \cite{Carrera:2009ve} which is incompatible with a canonical vector field and scalar in spherical symmetry.

To tackle the problem, one could try a reversed approach, starting from a known scalar-tensor black-hole solution that is close enough to the McVittie metric, or a generalization of it, and perturb it to try to get at least in some limit (for instance, at late times) a connection to metric \eqref{mcvittie}. One such example of a hairy black hole in a cosmological background is the so-called Martínez-Troncoso-Zanelli (MTZ) solution \cite{Martinez:2002ru,*Martinez:2004nb,Martinez:2005di}, which we consider as a starting point for our backward search.

To better understand the logic behind the perturbative approach we develop later, we first retrace the steps towards the MTZ solution and present a preliminary result. The MTZ metric is sourced by a conformally coupled scalar and a vector field, but for simplicity we begin by neglecting the vector and considering the system described by the action
\begin{multline}
  I = \int \ud^4 x \sqrt{-g} \left[ \frac{R - 2 \Lambda}{16 \pi} - \frac{1}{2} g^{\alpha\beta} \phi_{;\alpha} \phi_{;\beta} \right.\\
  \left.- \frac{1}{12} R \phi^2 - \alpha \phi^4 \right] \,, \label{action}
\end{multline}
where $\alpha$ is a dimensionless constant. The corresponding field equations are
\begin{align} \label{fieldeqs}
  G_{\mu\nu} + \Lambda g_{\mu\nu} =&\, 8 \pi T_{\mu\nu}^{\text{(S)}} \,, \\
  \Box \phi =&\, \frac{1}{6} R \phi + 4 \alpha \phi^3 \,,\label{fieldeqs1}
\end{align}
where the energy-momentum tensor of the scalar field is
\begin{equation}\label{scalmom}
\begin{split}
  T_{\mu\nu}^{\text{(S)}} =&\, \partial_\mu \phi \partial_\nu \phi - \frac{1}{2} g_{\mu\nu} g^{\alpha\beta} \partial_\alpha \phi \partial_\beta \phi\\
  &\,+ \frac{1}{6} \left[ g_{\mu\nu} \Box - \nabla_\mu \nabla_\nu + G_{\mu\nu} \right] \phi^2 - \alpha g_{\mu\nu} \phi^4 \,,
\end{split}
\end{equation}
and $\Box \equiv g^{\alpha\beta} \nabla_{\alpha} \nabla_{\beta}$.

Since we look for spherically symmetric solutions with a cosmological constant, a possible heuristic approach is to write a line element of the form
\begin{equation} \label{eq1}
  \ud s^2 = - \left( 1 - \frac{m}{r} \right)^2 \ud t^2 + \frac{\ud r^2}{\left( 1 - \frac{m}{r} \right)^2} + r^2 \ud \Omega^2
\end{equation}
in isotropic coordinates, with $r = \rho + m$:
\begin{equation}\label{rho-r}
  \ud s^2 = -\! \left( \frac{\rho}{\rho + m} \right)^2 \!\ud t^2 \!+\! \left( 1 + \frac{m}{\rho} \right)^2 \!\! \left( \ud \rho^2 \!+\! \rho^2 \ud\Omega^2 \right) .
\end{equation}

If we take \eqref{rho-r} and replace $\rho (r)$ by a function of the form $\rho = \rho a (t)$, we can see that Eqs.\ \eqref{fieldeqs} and \eqref{fieldeqs1} are satisfied for
\begin{subequations}
\begin{gather}
  \begin{split}
    \ud s^2 =&\, - \left[ \frac{\rho}{\rho + \frac{m}{a (t)}} \right]^2 \ud t^2 \\
    &+ a (t)^2 \left[ 1 + \frac{m}{a (t) \rho} \right]^2 \left( \ud \rho^2 + \rho^2 \ud \Omega^2 \right) \,,\label{McVittieMTZ}
  \end{split}\\
  \intertext{with}
  \phi (t, \rho) = \sqrt{-\frac{\Lambda}{6 \alpha}} \frac{m}{a (t) \rho} \,,\label{McScalar}\\
  \intertext{and the scale factor is}
  a (t) = a_0 e^{\sqrt{\frac{\Lambda}{3}} t} \,.
\end{gather}
\end{subequations}
The solution given by \eqref{McVittieMTZ} and \eqref{McScalar} does not correspond, however, to a dynamical time-dependent black hole, since it can be rewritten as a stationary black hole by a coordinate transformation \cite{Brill:1993tw}.

To go further we need to allow more freedom by adding a vector field. Instead of trying to solve the resulting equations from the action \eqref{action} with a time-dependent scalar field and a generic spherically symmetric metric, we modify the action and consider
\begin{multline}
  I = \int \ud^4 x \sqrt{-g} \left[ \frac{R - 2 \Lambda}{16 \pi} - \frac{1}{2} g^{\mu\nu} \partial_\mu \phi \partial_\nu \phi \right. \\
  \left. - \frac{1}{12} R \phi^2 - \alpha \phi^4 - \frac{1}{16 \pi} F^{\mu\nu} F_{\mu\nu} \right] \,.\label{action1}
\end{multline}
An exact solution of the coupled Einstein-Maxwell-scalar field equations resulting from variation of the action \eqref{action1} is the MTZ solution cited before, and it reads
\begin{subequations} \label{sols}
  \begin{align}
    \begin{split}
      \ud s^2 =& - \left[ \left( 1 - \frac{m}{r} \right)^2 - \frac{\Lambda}{3} r^2 \right] \ud t^2 \\
      &+ \left[ \left( 1 - \frac{m}{r} \right)^2 - \frac{\Lambda}{3} r^2 \right]^{-1} \ud r^2 + r^2 \ud \Omega^2 \,,
    \end{split}\label{metric-mtz} \\
    \phi (r) =&\, \frac{m}{r - m} \sqrt{- \frac{\Lambda}{6 \alpha}} \,,\\
    A =&\, -\frac{q}{r} \ud t \,,
  \end{align}
\end{subequations}
with the extra condition
\begin{equation} \label{qcondition}
  q^2 = m^2 \left( 1 + \frac{2 \pi \Lambda}{9 \alpha} \right) \ .
\end{equation}
This solution describes a hairy static black-hole solution with a scalar field which is regular on the horizon and at infinity.

With the above results in mind, one can try to look for a generalization of the MTZ solution \eqref{sols} which allows for more general cosmological histories, while keeping the field content of the theory unchanged, namely, assuming \eqref{action1}. To introduce a general cosmological behavior we focus on the class of metrics that are obtained by adding a time dependence to what was the cosmological constant in \eqref{metric-mtz} sending $\Lambda \to 3 H(t)$, which corresponds to a generalization of Reissner-Nordstr\"{o}m-de~Sitter with a time-dependent Hubble factor. Even after making this specific choice for the metric, solving the equations directly is quite challenging, so to proceed we restrict our attention even further to the subclass of spacetimes that reduce asymptotically in time to a cosmological constant. This allows for a perturbative treatment at large $t$ that strongly simplifies the equations. Of course, the validity of the results will be limited to the class of systems that tend ``smoothly enough'' to the MTZ solution we have seen above, which was anyway our initial goal when we asked the question of whether or not the McVittie solution or some generalization thereof could be sourced by fields in an analogous way to the MTZ solution.

Starting then from the action \eqref{action1}, we look for solutions of the form
\begin{multline}
  \ud s^2 = - \left[ \left( 1 - \frac{m}{r} \right)^2 - H (t) r^2 \right] \ud t^2 \\
  + \left[ \left( 1 - \frac{m}{r} \right)^2 - H (t) r^2 \right]^{-1} \ud r^2 + r^2 \ud \Omega^2 \, , \label{mtz-McV}
\end{multline}
requiring in turn that $H(t) \overset{t \to \infty}{\to} \nicefrac{\Lambda}{3}$ with some inverse power of $t$, namely,
\begin{equation}\label{Hdecay}
H(t) \to \frac{H_1}{t^n} + H_0 \ , \qquad \forall n \ge 1 \ ,
\end{equation}
where $H_1$ is some positive constant and $H_0 = \nicefrac{\Lambda}{3}$. On the field side, the large-time corrections will be generally given by
\begin{align}
\phi(t,r) &= \phi_{\text{MTZ}}+\delta\phi(t,r) \,, \\
A_\mu(t,r) &= A^{\text{MTZ}}_\mu+ \delta A_\mu(t,r) \,,
\end{align}
where the $\text{MTZ}$ label refers to the solutions in Eqs.\ \eqref{sols} and the corrections die off at time infinity. 

With these assumptions one can expand both sides of the Einstein equations and, comparing the field perturbations to the decaying rate of the left-hand side defined by the form of $H(t)$ \eqref{Hdecay}, find the large-time behavior of $\delta\phi$ and $\delta A_\mu$. This turns out to be
\begin{align}
\delta\phi(t,r) &= \frac{\Phi(r)}{t^n} \\
\delta A_t(t,r) &= \frac{a_t (r)}{t^n} \\
\partial_t \delta A_r(t,r) &= \frac{a_r (r)}{t^n} \\
\delta A_\theta(t,r) &= \mathcal{O}\left(t^{-n-1}\right) \\
\delta A_\phi(t,r) &= \mathcal{O}\left(t^{-n-1}\right) \,.
\end{align}
Although these are the maximal perturbations allowed for the scalar and the vector, it is easy to see that allowing any of the above to be of higher order would eliminate from the game either the scalar or the vector, reducing even further the possibility to find solutions. In other words, $\delta \phi(t,r)$ could be taken of $\mathcal{O} \left( t^{-n - 1} \right)$, but this would just make all the scalar corrections drop from the equations at the order at which $H_1 t^{-n}$ corrections appear. Similarly, due to the fact that the corrections introduced in the metric respect spherical symmetry, of the four components of the vector field the angular ones must be negligible, and only the combination $\delta E_r (t,r) \equiv \partial_r A_t(t,r) - \partial_t A_r (t,r)$ is physical. Moreover, the latter is forced to decay as $t^{-n}$ for the same reasons discussed in the scalar field case.
 
Finally, one can search perturbatively for solutions of Einstein equations, only to find out that there is none. This result has to be read as the impossibility of perturbatively building a solution that smoothly tends to MTZ at large times [where smoothly is properly defined by the large time behavior $t^{-n}$ of $H (t)$, for all $n \ge 1$] starting with the ingredients in the action \eqref{action1}: namely, a conformally coupled scalar and a vector field, and the ansatz for the metric in \eqref{mtz-McV}.

\section{McVittie spacetime as a solution to a cosmological scalar field}\label{sec:cuscuton_all}

The results from Sec.\ \ref{sec:mtz-mcv} imply that canonical scalar or vector fields cannot act as sources to the generalized MTZ solution with an arbitrary cosmological history. If we consider the McVittie spacetime as the target solution, the restrictions become more severe, due to spatial Ricci isotropy of the metric \cite{Carrera:2009ve}, which can be physically interpreted as the fact that a comoving fluid must be devoid of anisotropic stresses in this spacetime. In this case, a canonical scalar (quintessence) field is automatically excluded as the sole source, since Ricci isotropy requires the field to be homogeneous. As a consequence, there can be no radial dependence on the potential, and therefore the characteristic inhomogeneous pressure which arises in McVittie cannot be obtained from the field. Similar inconsistencies arise when one also considers vector fields on the matter action.

A possible way out is to relax the assumptions on the scalar field, for instance, by allowing noncanonical kinetic terms in the action. In the following sections, we show that one such $k$-essence field can provide a consistent solution to the McVittie metric, both by direct inspection of the equations and from general requirements over the action.

\subsection{The $\mathbfit{k}$-essence field}

We begin by reviewing a few details of the $k$-essence field. Consider the scalar $\phi$ introduced \cite{ArmendarizPicon:2000ah,Babichev:2006vx,*Babichev:2007dw} via the action
\begin{equation}\label{actionessen}
  S_{\phi} = \int \ud^{4} x \sqrt{-g} \mathcal{L} \left( X, \phi \right) \,,
\end{equation}
where
\begin{equation}\label{eq:Xgen}
  X = -\frac{1}{2} g^{\mu\nu} \phi_{;\mu} \phi_{;\nu}
\end{equation}
is the canonical kinetic term of the field $\phi$. Variation of the action \eqref{actionessen} with respect to $\phi$ gives the $k$-essence equation of motion
\begin{equation}\label{EOM}
\begin{split}
  -\frac{\delta S}{\delta\phi} =&\, \left( \mathcal{L}_{,X} g^{\alpha\beta} + \mathcal{L}_{,XX} \phi^{;\alpha} \phi^{;\beta} \right) \phi_{;\alpha\beta}\\
  &\,+ 2 X \mathcal{L}_{,X \phi} - \mathcal{L}_{,\phi} \\
  =&\, 0 \,,
\end{split}
\end{equation}
so when the condition
\begin{equation}\label{HYPER}
  \mathcal{L}_{,X} + 2 X \mathcal{L}_{,XX} > 0 \,,
\end{equation}
holds, Eq.\ \eqref{EOM} is hyperbolic and $\phi$ describes a physical degree of freedom \cite{vikman-2007}.

Variation of the action \eqref{actionessen} with respect to $g_{\mu\nu}$ gives the energy-momentum tensor
\begin{equation}\label{EMT}
  T_{\mu\nu} \equiv \frac{2}{\sqrt{-g}} \frac{\delta S_{\phi}}{\delta g^{\mu\nu}} = \mathcal{L}_{,X} \phi_{;\mu} \phi_{;\nu} + g_{\mu\nu} \mathcal{L} \,.
\end{equation}
Note that the null energy condition also imposes $\mathcal{L}_{,X} \geq 0$. The form of the energy-momentum tensor defined by \eqref{EMT} allows us to define an equivalent fluid velocity \cite{vikman-2007}
\begin{equation}\label{eq:ueff}
  u_\mu \equiv -\frac{\nabla_\mu \phi}{\sqrt{2 X}} \,,
\end{equation}
so we can cast \eqref{EMT} in the form of a perfect fluid:
\begin{equation}\label{eq:tpf}
  T_{\mu\nu} = \rho\, u_\mu u_\nu + p\, h_{\mu\nu} \,,
\end{equation}
where $h_{\mu\nu} \equiv g_{\mu\nu} + u_{\mu} u_{\nu}$ is the projection tensor in the orthogonal direction to the flow, and the equivalent density and pressure are defined as
\begin{align}\label{eq:rhopeff}
  \rho &\equiv 2 X \mathcal{L}_{,X} -\mathcal{L},\\
  p &\equiv \mathcal{L}.
\end{align}

\subsection{Cuscuton field in spherical symmetry}

An interesting particular case of the wide family of $k$-essence systems is the cuscuton field \cite{Afshordi:2006ad,*Afshordi:2007yx}. This field is defined by choosing the following form for the action \eqref{actionessen}:
\begin{equation}\label{action-cus}
  S_{\phi} = \int \ud^4 x \sqrt{-g} \left[ \mu^2 \sqrt{-g^{\alpha\beta} \phi_{;\alpha} \phi_{;\beta}} - V (\phi) \right] \,,
\end{equation}
where $\mu$ is a constant. The equation of motion \eqref{EOM} then specializes to
\begin{equation}\label{eqmov-cus-gen}
  \frac{1}{\sqrt{-g}} \left( \frac{\sqrt{-g} \phi^{;\gamma}}{\sqrt{-g^{\alpha\beta} \phi_{;\alpha} \phi_{;\beta}}} \right)_{,\gamma} - \frac{1}{\mu^2} \frac{\ud V}{\ud \phi} = 0 \,,
\end{equation}
and the energy-momentum tensor \eqref{EMT} specializes to
\begin{equation}\label{temini-cusg}
\begin{split}
  T_{\mu\nu} =&\, \frac{\mu^2 \phi_{;\mu}\phi_{;\nu}}{\sqrt{-g^{\alpha\beta} \phi_{;\alpha} \phi_{;\beta}}}\\
  &\,+ g_{\mu\nu} \left[ \mu^2 \sqrt{-g^{\gamma\delta} \phi_{;\gamma} \phi_{;\delta}} - V (\phi) \right] \,.
\end{split}
\end{equation}

Assuming the most general spherically symmetric ansatz for the metric
\begin{equation}\label{sphsym}
  \ud s^2 = -e^{2 \nu (r, t)} \ud t^2 + e^{2 \lambda (r, t)} \ud r^2 + Y^2 (r, t) \ud \Omega^2 \,,
\end{equation}
and a homogeneous cuscuton field  $\phi = \varphi (t)$, the energy-momentum tensor \eqref{temini-cusg} reduces to
\begin{equation}\label{temini-cus}
  \tensor{T}{^\mu_\nu} = V (\varphi) u^{\mu} u_{\nu} + \left[ e^{-\nu} \mu^2 \dot{\varphi} - V (\varphi) \right] \tensor{h}{^\mu_\nu} \,.
\end{equation}
Comparison with Eq.~\eqref{eq:tpf} and homogeneity of the field allow us to associate the potential $V (\varphi)$, which is now just a function of time, with the energy density as
\begin{equation}\label{dens-cus}
  V (\varphi) = \rho (t) \,.
\end{equation}
The field equation \eqref{eqmov-cus-gen} then reduces to
\begin{equation}\label{eqmov-cus}
  \left( \frac{2 \dot{Y}}{Y} + \dot{\lambda} \right) e^{-\nu} = \frac{1}{\mu^2} \frac{\ud V}{\ud \varphi} \,.
\end{equation}
and using Eq.~\eqref{dens-cus} we can rewrite the right-hand side of Eq.~\eqref{eqmov-cus} as
\begin{equation}\label{BI}
  \frac{1}{\mu^2} \frac{\ud V}{\ud \varphi} = \frac{\dot{\rho}}{\mu^2 \dot{\varphi}}\,.
\end{equation}
Moreover, the expansion scalar $\Theta \equiv \tensor{u}{^{\mu}_{;\mu}}$ in terms of the comoving flow with respect to the line element \eqref{sphsym} reads
\begin{equation}\label{expans}
  \Theta = \left( \frac{2 \dot{Y}}{Y} + \dot{\lambda} \right) e^{-\nu}
\end{equation}
so Eq.~\eqref{eqmov-cus} really tells us that the expansion scalar is just a function of time. Although we started with the most general spherically symmetric metric, sourcing it with a homogeneous field with action given by \eqref{action-cus} simplifies things, and, in particular, one gets homogeneity in the density and expansion. We then write, as is customary in the case of homogeneous fluids, $\Theta \equiv 3 H(t)$. Inserting this definition for $H (t)$ into Eq.~\eqref{eqmov-cus}, we get
\begin{equation}\label{eqmov-cusH}
  3 H (t) = \frac{1}{\mu^2} \frac{\ud V}{\ud \varphi} \,.
\end{equation}
There are several equivalent ways to obtain the above result, be it from the conservation of the energy-momentum tensor \eqref{temini-cus}, whose only nonidentically satisfied component yields the same result as Eq.~\eqref{eqmov-cus}, or from the trace of the extrinsic curvature. More specifically, since the mean extrinsic curvature $\tensor{K}{^\alpha _\alpha}$ has the exact same expression as $\Theta$ from Eq.~\eqref{expans}, Eq.~\eqref{eqmov-cusH} gives
\begin{equation}\label{eq-cmc}
  \tensor{K}{^\alpha _\alpha} = \frac{1}{\mu^2} \frac{\ud V}{\ud \varphi} = 3 H(t) \,;
\end{equation}
that is, the mean curvature $\tensor{K}{^\alpha _\alpha}$ is constant along  the surfaces of constant $\phi$. 

It has been proven  \cite{Misra:1973zz} that a regular spherically symmetric perfect fluid with homogeneous energy density (in its comoving frame) can only drive uniform shear-free expansion (or contraction). We have shown in this section that the cuscuton action \eqref{action-cus} satisfies all the conditions necessary for this theorem to hold.

\subsection{McVittie as a solution to the cuscuton}\label{sec:mcv-cus}

There are two facts which motivate us to consider McVittie as a solution to a self-gravitating classical cuscuton. The first is the fact that the equivalent fluid description of McVittie metric possesses homogeneous density but inhomogeneous pressure, and, unlike a canonical scalar, a homogeneous cuscuton allows for homogeneous density without forcing a homogeneous pressure, as can be seen by inserting \eqref{dens-cus} in the first term of Eq.~\eqref{temini-cus}. The second comes from considering the extrinsic curvature in McVittie spacetime. By using the foliation defined by the comoving flow, the mean extrinsic curvature of metric \eqref{mcvittie} is
\begin{equation}
  \tensor{K}{^\alpha _\alpha} = 3\frac{\dot{a}}{a} \,,
\end{equation}
which is independent of the radial coordinate. Therefore, the line element \eqref{mcvittie} admits a foliation with constant mean curvature, which coincides with the comoving foliation used to obtain Eq.~\eqref{eq-cmc}.

Based on these similarities, we now assume the McVittie metric \eqref{mcvittie} as an ansatz for the system consisting of a self-gravitating cuscuton, and incidentally by comparison with Eq.~\eqref{eq-cmc} we make the familiar connection $\frac{\dot{a}}{a} = H$. In this metric, the Einstein equations for the cuscuton with the energy-momentum tensor given by \eqref{temini-cus} read
\begin{align}
  3 H^2 &= 8 \pi V (\varphi), \label{EEttMcV}\\
  2 \dot{H} \frac{2 a r + m}{2 a r - m} + 3 H^2 &= 8 \pi \left( V - \frac{2 a r + m}{2 a r - m} \mu^2 \dot{\varphi} \right)\,, \label{EErrMcV}
\end{align}
and the equation of motion \eqref{eqmov-cusH} reduces to
\begin{equation}\label{Eqmovcus}
  -3 \frac{\dot{a}}{a} + \frac{1}{\mu^2} \frac{\ud V}{\ud \varphi} = 0 \,.
\end{equation}

Substituting the potential from \eqref{EEttMcV} into \eqref{EErrMcV}, we find
\begin{equation}\label{eq:Hphidot}
4 \pi \mu^2 \dot{\varphi} = -\dot{H} \,.
\end{equation}

Using \eqref{EEttMcV} to eliminate $\frac{\dot{a}}{a}$ from \eqref{Eqmovcus}, we find
\begin{equation}
  \left( \frac{\ud V}{\ud \varphi} \right)^2 - 24 \pi \mu^4 V = 0 \,,
\end{equation}
which may be solved for $V$, so
\begin{equation}\label{eq:vcusmcv}
  V (\varphi) = 6 \pi \mu^4 \left( \varphi + C \right)^2 \,,
\end{equation}
where $C$ is an irrelevant integration constant.

Inserting \eqref{eq:Hphidot} into \eqref{EErrMcV} and \eqref{dens-cus} into \eqref{EEttMcV}, we recover the familiar forms for the density and pressure in the McVittie metric, and the system closes with the solution \eqref{eq:vcusmcv}. We can conclude therefore that, in fact, the McVittie metric is a consistent exact solution to the complete system consisting of a self-gravitating cuscuton minimally coupled to gravity.

\subsection{McVittie for a general $\mathbfit{k}$-essence}

Given the results of the previous section, we might now ask what the most general $k$-essence field that satisfies Einstein equations for McVittie spacetime \eqref{mcvittie} is. In other words, we now address the question of whether or not a homogeneous cuscuton obtained from the action \eqref{action-cus} is the unique $k$-essence field which is consistent with McVittie. Starting from the action \eqref{actionessen}, we may now cast the energy-momentum tensor \eqref{EMT} under metric \eqref{mcvittie} by using the equivalent fluid description.

So, in McVittie, the kinetic term $X$ from \eqref{eq:Xgen} reads
\begin{equation}\label{eq:Xcus}
  X = -\frac{8 a^2 r^4}{(2 a r + m)^4} (\phi\linha)^2 + \frac{1}{2}\frac{(2 a r + m)^2}{(2 a r - m)^2} \dot\phi^2 \,.
\end{equation}
Considering the equivalent four-velocity defined in \eqref{eq:ueff}, it becomes clear that, to satisfy the Einstein equations for McVittie, which requires a comoving perfect fluid, that is, that the off-diagonal terms of the energy-momentum tensor vanish, one must have either a constant field (that simultaneously leads to $\mathcal{L}_{,X} = 0$) or $\phi = \varphi (t)$. In other words, a homogeneous $k$-essence field is a necessary condition for McVittie to be a solution. A uniform field immediately implies that the spatial Ricci-isotropy condition is satisfied, another requirement of the McVittie solution \cite{Carrera:2009ve}, so the remaining independent equations are the $(t, t)$ and the $(r, r)$ components of Einstein's equations and the equation of motion for the $k$-essence field.

We have seen in Sec.~\ref{sec:mcv-cus} that the special case in which  $\mathcal{L} (X, \varphi) = \mu^2 \sqrt{X} - V (\varphi)$ is a solution, so now we want to constrain the functional form of possible $\mathcal{L}$. In order to do so we perform a simple check: considering the $tt$ component of Einstein's equations
\begin{equation}\label{EE00k}
  -2 X \mathcal{L}_{,X} + \mathcal{L} = -\frac{3}{8 \pi} \left( \frac{\dot{a}}{a} \right)^2 \,,
\end{equation}
we realize that the right-hand side does not contain any radial coordinate dependence. We can then take its derivative with respect to $r$, to obtain the constraint equation
\begin{equation}
4 m a \dot{\varphi}^2 \frac{(2 a r + m)}{(2 a r - m)^3} \left( 2 X \mathcal{L}_{,XX} + \mathcal{L}_{,X} \right)= 0 \,,
\end{equation}
which vanishes only if the last term vanishes. We can then formally integrate it and find the solution
\begin{equation}\label{Cuscuton-only}
  \mathcal{L} (X, \varphi) = A (\varphi) + B (\varphi) \sqrt{X} \,.
\end{equation}
This proves that the cuscuton is the only $k$-essence model coming from action \eqref{actionessen} that can support the McVittie geometry, given that the integration constant $A$ in \eqref{Cuscuton-only} can be reabsorbed in the potential $V(\varphi)$ and that $B$ can be reabsorbed with a trivial field redefinition $\varphi \to \phi = \mu^{-2}\int  \sqrt{B(\varphi)} d\varphi$. Note that \eqref{Cuscuton-only} violates the inequality \eqref{HYPER}, which means that the field $\varphi$ does not carry a physical degree of freedom but rather ``latches on'' the evolution of the gravitational part, itself fully determined by the shape of the potential as per Eq.~\eqref{EEttMcV}. This is a signature of the parasitic nature of the cuscuton field \cite{Afshordi:2006ad,*Afshordi:2007yx}.

\subsection{Ho\texorpdfstring{\v{r}}{ř}ava-Lifshitz gravity with anisotropic Weyl symmetry}

In this section, we make a surprising observation that connects McVittie cosmological black holes to a proposal for quantum gravity. Ho\v{r}ava-Lifshitz gravity is a proposal for a power-counting renormalizable theory of quantum gravity that trades Lorentz symmetry with $z=3$ Lifshitz symmetry (or anisotropic space and time scaling symmetry) at high energies \cite{horava-prd-2009}. However, effective field theory arguments suggest that relevant Lorentz-violating terms can be generated from quantum corrections at low energies and have been the subject of controversy and intense scrutiny. This relevant deformation is parametrized by the parameter $\lambda$, which is equal to $1$ for Einstein gravity, but its value is not protected by any symmetry in the Lorentz-violating theory. 

It turns out that the cuscuton action \eqref{action-cus} with a quadratic potential, coupled to Einstein gravity, is equivalent to the low-energy limit of the (nonprojectable) Ho\v{r}ava-Lifshitz gravity \cite{Afshordi:2009tt}. For a general quadratic potential $V(\phi) = \frac{1}{2} M^2\phi^2$, the Lorentz-violating parameter $\lambda$ in Ho\v{r}ava-Lifshitz gravity is given by
\begin{align}
  \lambda &= 1 - \frac{8 \pi \mu^4}{M^2} \nonumber\\
  &= 1- \frac{8\pi}{12\pi} = 1-\frac{2}{3} = \frac{1}{3},
\end{align}
where, in the second line, we used $M^2= 12\pi\mu^4$ from Eq.~\eqref{eq:vcusmcv} for the cuscuton theory that satisfies the McVittie geometry. 

The significance of $\lambda = \nicefrac{1}{3}$ was already pointed out by Ho\v{r}ava in his original paper \cite{horava-prd-2009}. While the $z=3$ Lifshitz \emph{global} anisotropic scaling is the sufficient condition for power-counting renormalizability of geometrodynamics, one may impose the stricter constraint of \emph{local} anisotropic Weyl symmetry, which significantly limits the allowed terms in the action\footnote{Furthermore, one can establish a formal (but nonlocal) equivalence between anisotropic Weyl+$3d$ diffeomorphisms and $4d$ diffeomorphisms, implying that the theory maintains the same number of degrees of freedom as general relativity. This forms the basis for the \emph{shape dynamics} approach to geometrodynamics \cite{Gomes:2010fh}.} and fixes $\lambda=\nicefrac{1}{3}$. The other term that is allowed by the Weyl symmetry is quadratic in the Cotton tensor for spatial 3-metric and thus vanishes for McVittie geometry which has a conformally flat (or shear-free) spatial geometry. 

Therefore, we conclude that McVittie black-hole spacetimes are also exact solutions to the vacuum Ho\v{r}ava-Lifshitz gravity with anisotropic Weyl symmetry.

\section{Conclusions and Future Prospects}\label{sec:conclusions}

In this work, we have looked at the problem of self-gravitating classical fields, specifically searching for dynamical spherically symmetric solutions which can describe black holes in an expanding universe, by concentrating on the particular cases of the MTZ and McVittie metrics. We have shown that a generalization of the MTZ metric with an arbitrary cosmological history \emph{cannot} be obtained perturbatively through the use of canonical fields. In McVittie, the situation is even more restrictive, since the symmetries of the Einstein tensor prohibit canonical fields from being homogeneous (a required condition imposed by Ricci isotropy, meaning that the equivalent fluid is free from anisotropic stresses) while retaining the characteristic inhomogeneous pressure present in the McVittie metric. However, by allowing for a noncanonical kinetic term in the scalar action, we have shown that it is possible to obtain a consistent solution: the cuscuton theory is the unique form of scalar field action which gives the McVittie metric when coupled to gravity. 

Furthermore, using the equivalence between cuscuton and low energy Ho\v{r}ava-Lifshitz gravity, we showed that McVittie geometry is also an exact vacuum solution for the full nonprojectable Ho\v{r}ava-Lifshitz gravity with anisotropic Weyl symmetry.

This result opens the possibility of analyzing the properties of self-gravitating fields on richer backgrounds than Minkowski or Schwarzschild. Moreover, despite the fact that McVittie is unlikely to describe a realistic classical fluid configuration, the fact that it is a solution to a scalar field is a decisive step forward in studies of modified gravity, since many modified gravity theories can be restated in the Einstein frame as scalar degrees of freedom coupled to general relativity.

Let us make some final observations that may form the bases for directions of future inquiry.

\begin{enumerate}

\item If we consider the Ho\v{r}ava-Lifshitz theory with Weyl symmetry to be a viable UV completion of gravity, then McVittie will be an appropriate geometry for primordial black holes that existed at the moment of the big bang, prior to the Planck time. This potentially opens an unprecedented window into the inhomogeneous dynamics of the big bang.

\item It is interesting that the most general known exact solutions for cosmological black holes, LTB and McVittie spacetimes, are exact opposites, i.e., have fluids with zero and infinite speeds of sound, respectively. While LTB can have an arbitrary comoving density profile $\rho(r)$, McVittie has an arbitrary expansion history $a(t)$. Could these two spacetimes be connected via a (yet unknown) duality that replaces space and time?

\item Can McVittie black holes teach us something about the physics of black holes that we may encounter in nature? Furthermore, could McVittie provide a potential description for the interaction of dark energy with astrophysical black holes?

\end{enumerate}

These are all questions with various degrees of speculation and/or physical relevance that continue to extend the legacy of McVittie into the 21st century.

\begin{acknowledgments}

We thank E.\ Babichev, M.\ Le~Delliou, K.\ Hinterbichler and T.\ Kolyvaris for useful discussions. D.\ C.\ G.\ also thanks Perimeter Institute for hospitality. E.\ A.\ is supported by FAPESP and CNPq through grants and fellowships. M.\ F.\ is supported by FAPESP Grant No.~2011/11365-4. D.\ C.\ G.\ is supported by FAPESP Grants No.~2010/08267-8 and No.~2013/01854-3. This research was supported in part by Perimeter Institute for Theoretical Physics. Research at Perimeter Institute is supported by the Government of Canada through Industry Canada and by the Province of Ontario through the Ministry of Economic Development and Innovation.

\end{acknowledgments}

\bibliography{shortnames,referencias}

\newcommand{\grg}{Gen.\ Relativ.\ Gravit.\@}
\begin{thebibliography}{58}%
\makeatletter
\providecommand \@ifxundefined [1]{%
 \@ifx{#1\undefined}
}%
\providecommand \@ifnum [1]{%
 \ifnum #1\expandafter \@firstoftwo
 \else \expandafter \@secondoftwo
 \fi
}%
\providecommand \@ifx [1]{%
 \ifx #1\expandafter \@firstoftwo
 \else \expandafter \@secondoftwo
 \fi
}%
\providecommand \natexlab [1]{#1}%
\providecommand \enquote  [1]{``#1''}%
\providecommand \bibnamefont  [1]{#1}%
\providecommand \bibfnamefont [1]{#1}%
\providecommand \citenamefont [1]{#1}%
\providecommand \href@noop [0]{\@secondoftwo}%
\providecommand \href [0]{\begingroup \@sanitize@url \@href}%
\providecommand \@href[1]{\@@startlink{#1}\@@href}%
\providecommand \@@href[1]{\endgroup#1\@@endlink}%
\providecommand \@sanitize@url [0]{\catcode `\\12\catcode `\$12\catcode
  `\&12\catcode `\#12\catcode `\^12\catcode `\_12\catcode `\%12\relax}%
\providecommand \@@startlink[1]{}%
\providecommand \@@endlink[0]{}%
\providecommand \url  [0]{\begingroup\@sanitize@url \@url }%
\providecommand \@url [1]{\endgroup\@href {#1}{\urlprefix }}%
\providecommand \urlprefix  [0]{URL }%
\providecommand \Eprint [0]{\href }%
\providecommand \doibase [0]{http://dx.doi.org/}%
\providecommand \selectlanguage [0]{\@gobble}%
\providecommand \bibinfo  [0]{\@secondoftwo}%
\providecommand \bibfield  [0]{\@secondoftwo}%
\providecommand \translation [1]{[#1]}%
\providecommand \BibitemOpen [0]{}%
\providecommand \bibitemStop [0]{}%
\providecommand \bibitemNoStop [0]{.\EOS\space}%
\providecommand \EOS [0]{\spacefactor3000\relax}%
\providecommand \BibitemShut  [1]{\csname bibitem#1\endcsname}%
\let\auto@bib@innerbib\@empty
\bibitem [{\citenamefont {\relax{Yu}. Khlopov}\ \emph
  {et~al.}(1985)\citenamefont {\relax{Yu}. Khlopov}, \citenamefont {Malomed},\
  and\ \citenamefont {\relax{Ya}. B.~Zeldovich}}]{Khlopov:1985jw}%
  \BibitemOpen
  \bibfield  {author} {\bibinfo {author} {\bibfnamefont {M.}~\bibnamefont
  {\relax{Yu}. Khlopov}}, \bibinfo {author} {\bibfnamefont {B.}~\bibnamefont
  {Malomed}}, \ and\ \bibinfo {author} {\bibnamefont {\relax{Ya}.
  B.~Zeldovich}},\ }\href {http://adsabs.harvard.edu/abs/1985MNRAS.215..575K}
  {\bibfield  {journal} {\bibinfo  {journal} {Mon.\ Not.\ R.\ Astron.\ Soc.\@}\
  }\textbf {\bibinfo {volume} {215}},\ \bibinfo {pages} {575} (\bibinfo {year}
  {1985})}\BibitemShut {NoStop}%
\bibitem [{\citenamefont {Rubin}\ \emph {et~al.}(2001)\citenamefont {Rubin},
  \citenamefont {Sakharov},\ and\ \citenamefont {\relax{Yu}.
  Khlopov}}]{Rubin:2001yw}%
  \BibitemOpen
  \bibfield  {author} {\bibinfo {author} {\bibfnamefont {S.~G.}\ \bibnamefont
  {Rubin}}, \bibinfo {author} {\bibfnamefont {A.~S.}\ \bibnamefont {Sakharov}},
  \ and\ \bibinfo {author} {\bibfnamefont {M.}~\bibnamefont {\relax{Yu}.
  Khlopov}},\ }\href {\doibase 10.1134/1.1385631} {\bibfield  {journal}
  {\bibinfo  {journal} {J.\ Exp.\ Theor.\ Phys.\@}\ }\textbf {\bibinfo {volume}
  {92}},\ \bibinfo {pages} {921} (\bibinfo {year} {2001})},\ \Eprint
  {http://arxiv.org/abs/hep-ph/0106187} {arXiv:hep-ph/0106187} \BibitemShut
  {NoStop}%
\bibitem [{\citenamefont {\relax{Yu}. Khlopov}\ \emph
  {et~al.}(2002)\citenamefont {\relax{Yu}. Khlopov}, \citenamefont {Rubin},\
  and\ \citenamefont {Sakharov}}]{Khlopov:2002yi}%
  \BibitemOpen
  \bibfield  {author} {\bibinfo {author} {\bibfnamefont {M.}~\bibnamefont
  {\relax{Yu}. Khlopov}}, \bibinfo {author} {\bibfnamefont {S.~G.}\
  \bibnamefont {Rubin}}, \ and\ \bibinfo {author} {\bibfnamefont {A.~S.}\
  \bibnamefont {Sakharov}},\ }\href@noop {} {\bibfield  {journal} {\bibinfo
  {journal} {Gravitation Cosmol.\@}\ }\textbf {\bibinfo {volume} {8}},\
  \bibinfo {pages} {57} (\bibinfo {year} {2002})},\ \Eprint
  {http://arxiv.org/abs/astro-ph/0202505} {arXiv:astro-ph/0202505} \BibitemShut
  {NoStop}%
\bibitem [{\citenamefont {\relax{Yu}. Khlopov}\ \emph
  {et~al.}(2005)\citenamefont {\relax{Yu}. Khlopov}, \citenamefont {Rubin},\
  and\ \citenamefont {Sakharov}}]{Khlopov:2004sc}%
  \BibitemOpen
  \bibfield  {author} {\bibinfo {author} {\bibfnamefont {M.}~\bibnamefont
  {\relax{Yu}. Khlopov}}, \bibinfo {author} {\bibfnamefont {S.~G.}\
  \bibnamefont {Rubin}}, \ and\ \bibinfo {author} {\bibfnamefont {A.~S.}\
  \bibnamefont {Sakharov}},\ }\href {\doibase
  10.1016/j.astropartphys.2004.12.002} {\bibfield  {journal} {\bibinfo
  {journal} {Astropart.\ Phys.\@}\ }\textbf {\bibinfo {volume} {23}},\ \bibinfo
  {pages} {265} (\bibinfo {year} {2005})},\ \Eprint
  {http://arxiv.org/abs/astro-ph/0401532} {arXiv:astro-ph/0401532} \BibitemShut
  {NoStop}%
\bibitem [{\citenamefont {Maeda}\ and\ \citenamefont
  {Nozawa}(2010{\natexlab{a}})}]{Maeda:2009ds}%
  \BibitemOpen
  \bibfield  {author} {\bibinfo {author} {\bibfnamefont {K.-i.}\ \bibnamefont
  {Maeda}}\ and\ \bibinfo {author} {\bibfnamefont {M.}~\bibnamefont {Nozawa}},\
  }\href {\doibase 10.1103/PhysRevD.81.044017} {\bibfield  {journal} {\bibinfo
  {journal} {Phys.\ Rev.\ D}\ }\textbf {\bibinfo {volume} {81}},\ \bibinfo
  {pages} {044017} (\bibinfo {year} {2010}{\natexlab{a}})},\ \Eprint
  {http://arxiv.org/abs/0912.2811} {arXiv:0912.2811 [hep-th]} \BibitemShut
  {NoStop}%
\bibitem [{\citenamefont {Maeda}\ and\ \citenamefont
  {Nozawa}(2010{\natexlab{b}})}]{Maeda:2010ja}%
  \BibitemOpen
  \bibfield  {author} {\bibinfo {author} {\bibfnamefont {K.-i.}\ \bibnamefont
  {Maeda}}\ and\ \bibinfo {author} {\bibfnamefont {M.}~\bibnamefont {Nozawa}},\
  }\href {\doibase 10.1103/PhysRevD.81.124038} {\bibfield  {journal} {\bibinfo
  {journal} {Phys.\ Rev.\ D}\ }\textbf {\bibinfo {volume} {81}},\ \bibinfo
  {pages} {124038} (\bibinfo {year} {2010}{\natexlab{b}})},\ \Eprint
  {http://arxiv.org/abs/1003.2849} {arXiv:1003.2849 [gr-qc]} \BibitemShut
  {NoStop}%
\bibitem [{\citenamefont {Maeda}\ and\ \citenamefont
  {Nozawa}(2011)}]{Maeda:2011sh}%
  \BibitemOpen
  \bibfield  {author} {\bibinfo {author} {\bibfnamefont {K.-i.}\ \bibnamefont
  {Maeda}}\ and\ \bibinfo {author} {\bibfnamefont {M.}~\bibnamefont {Nozawa}},\
  }\href {\doibase 10.1143/PTPS.189.310} {\bibfield  {journal} {\bibinfo
  {journal} {Prog.\ Theor.\ Phys.\ Suppl.\@}\ }\textbf {\bibinfo {volume}
  {189}},\ \bibinfo {pages} {310} (\bibinfo {year} {2011})},\ \Eprint
  {http://arxiv.org/abs/1104.1849} {arXiv:1104.1849 [hep-th]} \BibitemShut
  {NoStop}%
\bibitem [{\citenamefont {Nozawa}\ and\ \citenamefont
  {Maeda}(2011)}]{Nozawa:2010zg}%
  \BibitemOpen
  \bibfield  {author} {\bibinfo {author} {\bibfnamefont {M.}~\bibnamefont
  {Nozawa}}\ and\ \bibinfo {author} {\bibfnamefont {K.-i.}\ \bibnamefont
  {Maeda}},\ }\href {\doibase 10.1103/PhysRevD.83.024018} {\bibfield  {journal}
  {\bibinfo  {journal} {Phys.\ Rev.\ D}\ }\textbf {\bibinfo {volume} {83}},\
  \bibinfo {pages} {024018} (\bibinfo {year} {2011})},\ \Eprint
  {http://arxiv.org/abs/1009.3688} {arXiv:1009.3688 [hep-th]} \BibitemShut
  {NoStop}%
\bibitem [{\citenamefont {Gibbons}\ and\ \citenamefont
  {Maeda}(2010)}]{Gibbons:2009dr}%
  \BibitemOpen
  \bibfield  {author} {\bibinfo {author} {\bibfnamefont {G.~W.}\ \bibnamefont
  {Gibbons}}\ and\ \bibinfo {author} {\bibfnamefont {K.-i.}\ \bibnamefont
  {Maeda}},\ }\href {\doibase 10.1103/PhysRevLett.104.131101} {\bibfield
  {journal} {\bibinfo  {journal} {Phys.\ Rev.\ Lett.\@}\ }\textbf {\bibinfo
  {volume} {104}},\ \bibinfo {pages} {131101} (\bibinfo {year} {2010})},\
  \Eprint {http://arxiv.org/abs/0912.2809} {arXiv:0912.2809 [gr-qc]}
  \BibitemShut {NoStop}%
\bibitem [{\citenamefont {Bocharova}\ \emph {et~al.}(1970)\citenamefont
  {Bocharova}, \citenamefont {Bronnikov},\ and\ \citenamefont
  {Mel'Nikov}}]{bocharova-1970}%
  \BibitemOpen
  \bibfield  {author} {\bibinfo {author} {\bibfnamefont {N.~M.}\ \bibnamefont
  {Bocharova}}, \bibinfo {author} {\bibfnamefont {K.~A.}\ \bibnamefont
  {Bronnikov}}, \ and\ \bibinfo {author} {\bibfnamefont {V.~N.}\ \bibnamefont
  {Mel'Nikov}},\ }\href {http://adsabs.harvard.edu/abs/1970VeMos..11..706B}
  {\bibfield  {journal} {\bibinfo  {journal} {Vestn.\ Mosk.\ Un-ta.\ Fiz.,
  Astron.\@}\ }\textbf {\bibinfo {volume} {11}},\ \bibinfo {pages} {706}
  (\bibinfo {year} {1970})}\BibitemShut {NoStop}%
\bibitem [{\citenamefont {Bekenstein}(1974)}]{bekenstein-1974}%
  \BibitemOpen
  \bibfield  {author} {\bibinfo {author} {\bibfnamefont {J.~D.}\ \bibnamefont
  {Bekenstein}},\ }\href {\doibase 10.1016/0003-4916(74)90124-9} {\bibfield
  {journal} {\bibinfo  {journal} {Ann.\ Phys.\ (N.Y.)}\ }\textbf {\bibinfo
  {volume} {82}},\ \bibinfo {pages} {535} (\bibinfo {year} {1974})}\BibitemShut
  {NoStop}%
\bibitem [{\citenamefont {Bekenstein}(1975)}]{bekenstein-1975}%
  \BibitemOpen
  \bibfield  {author} {\bibinfo {author} {\bibfnamefont {J.~D.}\ \bibnamefont
  {Bekenstein}},\ }\href {\doibase 10.1016/0003-4916(75)90279-1} {\bibfield
  {journal} {\bibinfo  {journal} {Ann.\ Phys.\ (N.Y.)}\ }\textbf {\bibinfo
  {volume} {91}},\ \bibinfo {pages} {75} (\bibinfo {year} {1975})}\BibitemShut
  {NoStop}%
\bibitem [{\citenamefont {Bronnikov}\ and\ \citenamefont {\relax{Yu}.
  N.~Kireyev}(1978)}]{bronnikov-1978}%
  \BibitemOpen
  \bibfield  {author} {\bibinfo {author} {\bibfnamefont {K.~A.}\ \bibnamefont
  {Bronnikov}}\ and\ \bibinfo {author} {\bibnamefont {\relax{Yu}.
  N.~Kireyev}},\ }\href {\doibase 10.1016/0375-9601(78)90030-0} {\bibfield
  {journal} {\bibinfo  {journal} {Phys.\ Lett.\ A}\ }\textbf {\bibinfo {volume}
  {67}},\ \bibinfo {pages} {95 } (\bibinfo {year} {1978})}\BibitemShut
  {NoStop}%
\bibitem [{\citenamefont {Zloshchastiev}(2005)}]{Zloshchastiev:2004ny}%
  \BibitemOpen
  \bibfield  {author} {\bibinfo {author} {\bibfnamefont {K.~G.}\ \bibnamefont
  {Zloshchastiev}},\ }\href {\doibase 10.1103/PhysRevLett.94.121101} {\bibfield
   {journal} {\bibinfo  {journal} {Phys.\ Rev.\ Lett.\@}\ }\textbf {\bibinfo
  {volume} {94}},\ \bibinfo {pages} {121101} (\bibinfo {year} {2005})},\
  \Eprint {http://arxiv.org/abs/hep-th/0408163} {arXiv:hep-th/0408163}
  \BibitemShut {NoStop}%
\bibitem [{\citenamefont {Mart\'{\i}nez}\ \emph {et~al.}(2003)\citenamefont
  {Mart\'{\i}nez}, \citenamefont {Troncoso},\ and\ \citenamefont
  {Zanelli}}]{Martinez:2002ru}%
  \BibitemOpen
  \bibfield  {author} {\bibinfo {author} {\bibfnamefont {C.}~\bibnamefont
  {Mart\'{\i}nez}}, \bibinfo {author} {\bibfnamefont {R.}~\bibnamefont
  {Troncoso}}, \ and\ \bibinfo {author} {\bibfnamefont {J.}~\bibnamefont
  {Zanelli}},\ }\href {\doibase 10.1103/PhysRevD.67.024008} {\bibfield
  {journal} {\bibinfo  {journal} {Phys.\ Rev.\ D}\ }\textbf {\bibinfo {volume}
  {67}},\ \bibinfo {pages} {024008} (\bibinfo {year} {2003})},\ \Eprint
  {http://arxiv.org/abs/hep-th/0205319} {arXiv:hep-th/0205319} \BibitemShut
  {NoStop}%
\bibitem [{\citenamefont {Mart\'{\i}nez}\ \emph {et~al.}(2004)\citenamefont
  {Mart\'{\i}nez}, \citenamefont {Troncoso},\ and\ \citenamefont
  {Zanelli}}]{Martinez:2004nb}%
  \BibitemOpen
  \bibfield  {author} {\bibinfo {author} {\bibfnamefont {C.}~\bibnamefont
  {Mart\'{\i}nez}}, \bibinfo {author} {\bibfnamefont {R.}~\bibnamefont
  {Troncoso}}, \ and\ \bibinfo {author} {\bibfnamefont {J.}~\bibnamefont
  {Zanelli}},\ }\href {\doibase 10.1103/PhysRevD.70.084035} {\bibfield
  {journal} {\bibinfo  {journal} {Phys.\ Rev.\ D}\ }\textbf {\bibinfo {volume}
  {70}},\ \bibinfo {pages} {084035} (\bibinfo {year} {2004})},\ \Eprint
  {http://arxiv.org/abs/hep-th/0406111} {arXiv:hep-th/0406111} \BibitemShut
  {NoStop}%
\bibitem [{\citenamefont {Sushkov}(2009)}]{Sushkov:2009hk}%
  \BibitemOpen
  \bibfield  {author} {\bibinfo {author} {\bibfnamefont {S.~V.}\ \bibnamefont
  {Sushkov}},\ }\href {\doibase 10.1103/PhysRevD.80.103505} {\bibfield
  {journal} {\bibinfo  {journal} {Phys.\ Rev.\ D}\ }\textbf {\bibinfo {volume}
  {80}},\ \bibinfo {pages} {103505} (\bibinfo {year} {2009})},\ \Eprint
  {http://arxiv.org/abs/0910.0980} {arXiv:0910.0980 [gr-qc]} \BibitemShut
  {NoStop}%
\bibitem [{\citenamefont {Kolyvaris}\ \emph {et~al.}(2012)\citenamefont
  {Kolyvaris}, \citenamefont {Koutsoumbas}, \citenamefont {Papantonopoulos},\
  and\ \citenamefont {Siopsis}}]{Kolyvaris:2011fk}%
  \BibitemOpen
  \bibfield  {author} {\bibinfo {author} {\bibfnamefont {T.}~\bibnamefont
  {Kolyvaris}}, \bibinfo {author} {\bibfnamefont {G.}~\bibnamefont
  {Koutsoumbas}}, \bibinfo {author} {\bibfnamefont {E.}~\bibnamefont
  {Papantonopoulos}}, \ and\ \bibinfo {author} {\bibfnamefont {G.}~\bibnamefont
  {Siopsis}},\ }\href {\doibase 10.1088/0264-9381/29/20/205011} {\bibfield
  {journal} {\bibinfo  {journal} {Classical Quantum Gravity}\ }\textbf
  {\bibinfo {volume} {29}},\ \bibinfo {pages} {205011} (\bibinfo {year}
  {2012})},\ \Eprint {http://arxiv.org/abs/1111.0263} {arXiv:1111.0263 [gr-qc]}
  \BibitemShut {NoStop}%
\bibitem [{\citenamefont {Kolyvaris}\ \emph {et~al.}(2013)\citenamefont
  {Kolyvaris}, \citenamefont {Koutsoumbas}, \citenamefont {Papantonopoulos},\
  and\ \citenamefont {Siopsis}}]{Kolyvaris:2013zfa}%
  \BibitemOpen
  \bibfield  {author} {\bibinfo {author} {\bibfnamefont {T.}~\bibnamefont
  {Kolyvaris}}, \bibinfo {author} {\bibfnamefont {G.}~\bibnamefont
  {Koutsoumbas}}, \bibinfo {author} {\bibfnamefont {E.}~\bibnamefont
  {Papantonopoulos}}, \ and\ \bibinfo {author} {\bibfnamefont {G.}~\bibnamefont
  {Siopsis}},\ }\href {\doibase 10.1007/JHEP11(2013)133} {\bibfield  {journal}
  {\bibinfo  {journal} {J.\ High Energy Phys.\@}\ }\textbf {\bibinfo {volume}
  {2013}},\ \bibinfo {pages} {1} (\bibinfo {year} {2013})},\ \Eprint
  {http://arxiv.org/abs/1308.5280} {arXiv:1308.5280 [hep-th]} \BibitemShut
  {NoStop}%
\bibitem [{\citenamefont {Fujii}(1982)}]{Fujii:1982ms}%
  \BibitemOpen
  \bibfield  {author} {\bibinfo {author} {\bibfnamefont {Y.}~\bibnamefont
  {Fujii}},\ }\href {\doibase 10.1103/PhysRevD.26.2580} {\bibfield  {journal}
  {\bibinfo  {journal} {Phys.\ Rev.\ D}\ }\textbf {\bibinfo {volume} {26}},\
  \bibinfo {pages} {2580} (\bibinfo {year} {1982})}\BibitemShut {NoStop}%
\bibitem [{\citenamefont {Ford}(1987)}]{Ford:1987de}%
  \BibitemOpen
  \bibfield  {author} {\bibinfo {author} {\bibfnamefont {L.~H.}\ \bibnamefont
  {Ford}},\ }\href {\doibase 10.1103/PhysRevD.35.2339} {\bibfield  {journal}
  {\bibinfo  {journal} {Phys.\ Rev.\ D}\ }\textbf {\bibinfo {volume} {35}},\
  \bibinfo {pages} {2339} (\bibinfo {year} {1987})}\BibitemShut {NoStop}%
\bibitem [{\citenamefont {Wetterich}(1988)}]{Wetterich:1987fm}%
  \BibitemOpen
  \bibfield  {author} {\bibinfo {author} {\bibfnamefont {C.}~\bibnamefont
  {Wetterich}},\ }\href {\doibase 10.1016/0550-3213(88)90193-9} {\bibfield
  {journal} {\bibinfo  {journal} {Nucl.\ Phys. B}\ }\textbf {\bibinfo {volume}
  {302}},\ \bibinfo {pages} {668} (\bibinfo {year} {1988})}\BibitemShut
  {NoStop}%
\bibitem [{\citenamefont {Armendariz-Picon}\ \emph {et~al.}(2001)\citenamefont
  {Armendariz-Picon}, \citenamefont {Mukhanov},\ and\ \citenamefont
  {Steinhardt}}]{ArmendarizPicon:2000ah}%
  \BibitemOpen
  \bibfield  {author} {\bibinfo {author} {\bibfnamefont {C.}~\bibnamefont
  {Armendariz-Picon}}, \bibinfo {author} {\bibfnamefont {V.}~\bibnamefont
  {Mukhanov}}, \ and\ \bibinfo {author} {\bibfnamefont {P.~J.}\ \bibnamefont
  {Steinhardt}},\ }\href {\doibase 10.1103/PhysRevD.63.103510} {\bibfield
  {journal} {\bibinfo  {journal} {Phys.\ Rev.\ D}\ }\textbf {\bibinfo {volume}
  {63}},\ \bibinfo {pages} {103510} (\bibinfo {year} {2001})},\ \Eprint
  {http://arxiv.org/abs/astro-ph/0006373} {arXiv:astro-ph/0006373} \BibitemShut
  {NoStop}%
\bibitem [{\citenamefont {Vaidya}(1951)}]{Vaidya:1951zz}%
  \BibitemOpen
  \bibfield  {author} {\bibinfo {author} {\bibfnamefont {P.~C.}\ \bibnamefont
  {Vaidya}},\ }\href@noop {} {\bibfield  {journal} {\bibinfo  {journal} {Proc.\
  Indian Acad.\ Sci.\ A}\ }\textbf {\bibinfo {volume} {33}},\ \bibinfo {pages}
  {264} (\bibinfo {year} {1951})},\ \bibinfo {note} {reprinted in
  \href{http://dx.doi.org/10.1023/A\%3A1018875606950}{\grg{} \textbf{31}(1),
  121--135 (1999)}}\BibitemShut {NoStop}%
\bibitem [{\citenamefont {Podolsk\'{y}}\ and\ \citenamefont
  {Sv\'{\i}tek}(2005)}]{Podolsky:2005ns}%
  \BibitemOpen
  \bibfield  {author} {\bibinfo {author} {\bibfnamefont {J.}~\bibnamefont
  {Podolsk\'{y}}}\ and\ \bibinfo {author} {\bibfnamefont {O.}~\bibnamefont
  {Sv\'{\i}tek}},\ }\href {\doibase 10.1103/PhysRevD.71.124001} {\bibfield
  {journal} {\bibinfo  {journal} {Phys.\ Rev.\ D}\ }\textbf {\bibinfo {volume}
  {71}},\ \bibinfo {pages} {124001} (\bibinfo {year} {2005})},\ \Eprint
  {http://arxiv.org/abs/gr-qc/0506016} {arXiv:gr-qc/0506016} \BibitemShut
  {NoStop}%
\bibitem [{\citenamefont {Lema\^{\i}tre}(1933)}]{Lemaitre:1933gd}%
  \BibitemOpen
  \bibfield  {author} {\bibinfo {author} {\bibfnamefont {G.}~\bibnamefont
  {Lema\^{\i}tre}},\ }\href@noop {} {\bibfield  {journal} {\bibinfo  {journal}
  {Annales Soc.\ Sci.\ Brux.\ Ser.\ A}\ }\textbf {\bibinfo {volume} {53}},\
  \bibinfo {pages} {51} (\bibinfo {year} {1933})},\ \bibinfo {note} {reprinted
  in \href{http://dx.doi.org/10.1023/A:1018855621348}{\grg{} \textbf{29}(5),
  641--680 (1997)}}\BibitemShut {NoStop}%
\bibitem [{\citenamefont {Tolman}(1934)}]{Tolman:1934za}%
  \BibitemOpen
  \bibfield  {author} {\bibinfo {author} {\bibfnamefont {R.~C.}\ \bibnamefont
  {Tolman}},\ }\href {\doibase 10.1073/pnas.20.3.169} {\bibfield  {journal}
  {\bibinfo  {journal} {Proc.\ Natl.\ Acad.\ Sci. U.S.A.\@}\ }\textbf {\bibinfo
  {volume} {20}},\ \bibinfo {pages} {169} (\bibinfo {year} {1934})}\BibitemShut
  {NoStop}%
\bibitem [{\citenamefont {Bondi}(1947)}]{bondi-1947}%
  \BibitemOpen
  \bibfield  {author} {\bibinfo {author} {\bibfnamefont {H.}~\bibnamefont
  {Bondi}},\ }\href {http://adsabs.harvard.edu/abs/1947MNRAS.107..410B}
  {\bibfield  {journal} {\bibinfo  {journal} {Mon.\ Not.\ R.\ Astron.\ Soc.\@}\
  }\textbf {\bibinfo {volume} {107}},\ \bibinfo {pages} {410} (\bibinfo {year}
  {1947})}\BibitemShut {NoStop}%
\bibitem [{\citenamefont {Firouzjaee}\ and\ \citenamefont
  {Mansouri}(2010)}]{firouzjaee-2010}%
  \BibitemOpen
  \bibfield  {author} {\bibinfo {author} {\bibfnamefont {J.~T.}\ \bibnamefont
  {Firouzjaee}}\ and\ \bibinfo {author} {\bibfnamefont {R.}~\bibnamefont
  {Mansouri}},\ }\href {\doibase 10.1007/s10714-010-0991-7} {\bibfield
  {journal} {\bibinfo  {journal} {\grg}\ }\textbf {\bibinfo {volume} {42}},\
  \bibinfo {pages} {2431} (\bibinfo {year} {2010})},\ \Eprint
  {http://arxiv.org/abs/0812.5108} {arXiv:0812.5108 [gr-qc]} \BibitemShut
  {NoStop}%
\bibitem [{\citenamefont {Firouzjaee}\ and\ \citenamefont
  {Mansouri}(2012)}]{Firouzjaee:2011hi}%
  \BibitemOpen
  \bibfield  {author} {\bibinfo {author} {\bibfnamefont {J.~T.}\ \bibnamefont
  {Firouzjaee}}\ and\ \bibinfo {author} {\bibfnamefont {R.}~\bibnamefont
  {Mansouri}},\ }\href {\doibase 10.1209/0295-5075/97/29002} {\bibfield
  {journal} {\bibinfo  {journal} {Europhys.\ Lett.\@}\ }\textbf {\bibinfo
  {volume} {97}},\ \bibinfo {pages} {29002} (\bibinfo {year} {2012})},\ \Eprint
  {http://arxiv.org/abs/1104.0530} {arXiv:1104.0530 [gr-qc]} \BibitemShut
  {NoStop}%
\bibitem [{\citenamefont {Einstein}\ and\ \citenamefont
  {Straus}(1945)}]{Einstein:1945id}%
  \BibitemOpen
  \bibfield  {author} {\bibinfo {author} {\bibfnamefont {A.}~\bibnamefont
  {Einstein}}\ and\ \bibinfo {author} {\bibfnamefont {E.~G.}\ \bibnamefont
  {Straus}},\ }\href {\doibase 10.1103/RevModPhys.17.120} {\bibfield  {journal}
  {\bibinfo  {journal} {Rev.\ Mod.\ Phys.\@}\ }\textbf {\bibinfo {volume}
  {17}},\ \bibinfo {pages} {120} (\bibinfo {year} {1945})}\BibitemShut
  {NoStop}%
\bibitem [{\citenamefont {Marra}\ \emph {et~al.}(2007)\citenamefont {Marra},
  \citenamefont {Kolb}, \citenamefont {Matarrese},\ and\ \citenamefont
  {Riotto}}]{Marra:2007pm}%
  \BibitemOpen
  \bibfield  {author} {\bibinfo {author} {\bibfnamefont {V.}~\bibnamefont
  {Marra}}, \bibinfo {author} {\bibfnamefont {E.~W.}\ \bibnamefont {Kolb}},
  \bibinfo {author} {\bibfnamefont {S.}~\bibnamefont {Matarrese}}, \ and\
  \bibinfo {author} {\bibfnamefont {A.}~\bibnamefont {Riotto}},\ }\href
  {\doibase 10.1103/PhysRevD.76.123004} {\bibfield  {journal} {\bibinfo
  {journal} {Phys.\ Rev.\ D}\ }\textbf {\bibinfo {volume} {76}},\ \bibinfo
  {pages} {123004} (\bibinfo {year} {2007})},\ \Eprint
  {http://arxiv.org/abs/0708.3622} {arXiv:0708.3622 [astro-ph]} \BibitemShut
  {NoStop}%
\bibitem [{\citenamefont {McVittie}(1933)}]{mcvittie-1933}%
  \BibitemOpen
  \bibfield  {author} {\bibinfo {author} {\bibfnamefont {G.~C.}\ \bibnamefont
  {McVittie}},\ }\href {http://adsabs.harvard.edu/abs/1933MNRAS..93..325M}
  {\bibfield  {journal} {\bibinfo  {journal} {Mon.\ Not.\ R.\ Astron.\ Soc.\@}\
  }\textbf {\bibinfo {volume} {93}},\ \bibinfo {pages} {325} (\bibinfo {year}
  {1933})}\BibitemShut {NoStop}%
\bibitem [{\citenamefont {Nolan}(1998)}]{Nolan:1998xs}%
  \BibitemOpen
  \bibfield  {author} {\bibinfo {author} {\bibfnamefont {B.~C.}\ \bibnamefont
  {Nolan}},\ }\href {\doibase 10.1103/PhysRevD.58.064006} {\bibfield  {journal}
  {\bibinfo  {journal} {Phys.\ Rev.\ D}\ }\textbf {\bibinfo {volume} {58}},\
  \bibinfo {pages} {064006} (\bibinfo {year} {1998})},\ \Eprint
  {http://arxiv.org/abs/gr-qc/9805041} {arXiv:gr-qc/9805041} \BibitemShut
  {NoStop}%
\bibitem [{\citenamefont {Nolan}(1999)}]{Nolan:1999kk}%
  \BibitemOpen
  \bibfield  {author} {\bibinfo {author} {\bibfnamefont {B.~C.}\ \bibnamefont
  {Nolan}},\ }\href {\doibase 10.1088/0264-9381/16/4/012} {\bibfield  {journal}
  {\bibinfo  {journal} {Classical Quantum Gravity}\ }\textbf {\bibinfo {volume}
  {16}},\ \bibinfo {pages} {1227} (\bibinfo {year} {1999})}\BibitemShut
  {NoStop}%
\bibitem [{\citenamefont {Kaloper}\ \emph {et~al.}(2010)\citenamefont
  {Kaloper}, \citenamefont {Kleban},\ and\ \citenamefont
  {Martin}}]{Kaloper:2010ec}%
  \BibitemOpen
  \bibfield  {author} {\bibinfo {author} {\bibfnamefont {N.}~\bibnamefont
  {Kaloper}}, \bibinfo {author} {\bibfnamefont {M.}~\bibnamefont {Kleban}}, \
  and\ \bibinfo {author} {\bibfnamefont {D.}~\bibnamefont {Martin}},\ }\href
  {\doibase 10.1103/PhysRevD.81.104044} {\bibfield  {journal} {\bibinfo
  {journal} {Phys.\ Rev.\ D}\ }\textbf {\bibinfo {volume} {81}},\ \bibinfo
  {pages} {104044} (\bibinfo {year} {2010})},\ \Eprint
  {http://arxiv.org/abs/1003.4777} {arXiv:1003.4777 [hep-th]} \BibitemShut
  {NoStop}%
\bibitem [{\citenamefont {Landry}\ \emph {et~al.}(2012)\citenamefont {Landry},
  \citenamefont {Abdelqader},\ and\ \citenamefont {Lake}}]{Landry:2012nv}%
  \BibitemOpen
  \bibfield  {author} {\bibinfo {author} {\bibfnamefont {P.}~\bibnamefont
  {Landry}}, \bibinfo {author} {\bibfnamefont {M.}~\bibnamefont {Abdelqader}},
  \ and\ \bibinfo {author} {\bibfnamefont {K.}~\bibnamefont {Lake}},\ }\href
  {\doibase 10.1103/PhysRevD.86.084002} {\bibfield  {journal} {\bibinfo
  {journal} {Phys.\ Rev.\ D}\ }\textbf {\bibinfo {volume} {86}},\ \bibinfo
  {pages} {084002} (\bibinfo {year} {2012})},\ \Eprint
  {http://arxiv.org/abs/1207.6350} {arXiv:1207.6350 [gr-qc]} \BibitemShut
  {NoStop}%
\bibitem [{\citenamefont {Jacobson}(1999)}]{Jacobson:1999vr}%
  \BibitemOpen
  \bibfield  {author} {\bibinfo {author} {\bibfnamefont {T.}~\bibnamefont
  {Jacobson}},\ }\href {\doibase 10.1103/PhysRevLett.83.2699} {\bibfield
  {journal} {\bibinfo  {journal} {Phys.\ Rev.\ Lett.\@}\ }\textbf {\bibinfo
  {volume} {83}},\ \bibinfo {pages} {2699} (\bibinfo {year} {1999})},\ \Eprint
  {http://arxiv.org/abs/astro-ph/9905303} {arXiv:astro-ph/9905303} \BibitemShut
  {NoStop}%
\bibitem [{\citenamefont {Frolov}\ and\ \citenamefont
  {Kofman}(2003)}]{Frolov:2002va}%
  \BibitemOpen
  \bibfield  {author} {\bibinfo {author} {\bibfnamefont {A.~V.}\ \bibnamefont
  {Frolov}}\ and\ \bibinfo {author} {\bibfnamefont {L.}~\bibnamefont
  {Kofman}},\ }\href {\doibase 10.1088/1475-7516/2003/05/009} {\bibfield
  {journal} {\bibinfo  {journal} {J.\ Cosmol.\ Astropart.\ Phys.\@}\ }\textbf
  {\bibinfo {volume} {2003}},\ \bibinfo {pages} {009} (\bibinfo {year}
  {2003})},\ \Eprint {http://arxiv.org/abs/hep-th/0212327}
  {arXiv:hep-th/0212327} \BibitemShut {NoStop}%
\bibitem [{\citenamefont {Saravani}\ \emph {et~al.}(2014)\citenamefont
  {Saravani}, \citenamefont {Afshordi},\ and\ \citenamefont
  {Mann}}]{Saravani:2013kva}%
  \BibitemOpen
  \bibfield  {author} {\bibinfo {author} {\bibfnamefont {M.}~\bibnamefont
  {Saravani}}, \bibinfo {author} {\bibfnamefont {N.}~\bibnamefont {Afshordi}},
  \ and\ \bibinfo {author} {\bibfnamefont {R.~B.}\ \bibnamefont {Mann}},\
  }\href {\doibase 10.1103/PhysRevD.89.084029} {\bibfield  {journal} {\bibinfo
  {journal} {Phys.\ Rev.\ D}\ }\textbf {\bibinfo {volume} {89}},\ \bibinfo
  {pages} {084029} (\bibinfo {year} {2014})},\ \Eprint
  {http://arxiv.org/abs/1310.4143} {arXiv:1310.4143 [gr-qc]} \BibitemShut
  {NoStop}%
\bibitem [{\citenamefont {Chadburn}\ and\ \citenamefont
  {Gregory}(2013)}]{Chadburn:2013mta}%
  \BibitemOpen
  \bibfield  {author} {\bibinfo {author} {\bibfnamefont {S.}~\bibnamefont
  {Chadburn}}\ and\ \bibinfo {author} {\bibfnamefont {R.}~\bibnamefont
  {Gregory}},\ }\href@noop {} {\  (\bibinfo {year} {2013})},\ \Eprint
  {http://arxiv.org/abs/1304.6287} {arXiv:1304.6287 [gr-qc]} \BibitemShut
  {NoStop}%
\bibitem [{\citenamefont {Afshordi}\ \emph
  {et~al.}(2007{\natexlab{a}})\citenamefont {Afshordi}, \citenamefont {Chung},\
  and\ \citenamefont {Geshnizjani}}]{Afshordi:2006ad}%
  \BibitemOpen
  \bibfield  {author} {\bibinfo {author} {\bibfnamefont {N.}~\bibnamefont
  {Afshordi}}, \bibinfo {author} {\bibfnamefont {D.~J.~H.}\ \bibnamefont
  {Chung}}, \ and\ \bibinfo {author} {\bibfnamefont {G.}~\bibnamefont
  {Geshnizjani}},\ }\href {\doibase 10.1103/PhysRevD.75.083513} {\bibfield
  {journal} {\bibinfo  {journal} {Phys.\ Rev.\ D}\ }\textbf {\bibinfo {volume}
  {75}},\ \bibinfo {pages} {083513} (\bibinfo {year} {2007}{\natexlab{a}})},\
  \Eprint {http://arxiv.org/abs/hep-th/0609150} {arXiv:hep-th/0609150}
  \BibitemShut {NoStop}%
\bibitem [{\citenamefont {Afshordi}\ \emph
  {et~al.}(2007{\natexlab{b}})\citenamefont {Afshordi}, \citenamefont {Chung},
  \citenamefont {Doran},\ and\ \citenamefont {Geshnizjani}}]{Afshordi:2007yx}%
  \BibitemOpen
  \bibfield  {author} {\bibinfo {author} {\bibfnamefont {N.}~\bibnamefont
  {Afshordi}}, \bibinfo {author} {\bibfnamefont {D.~J.~H.}\ \bibnamefont
  {Chung}}, \bibinfo {author} {\bibfnamefont {M.}~\bibnamefont {Doran}}, \ and\
  \bibinfo {author} {\bibfnamefont {G.}~\bibnamefont {Geshnizjani}},\ }\href
  {\doibase 10.1103/PhysRevD.75.123509} {\bibfield  {journal} {\bibinfo
  {journal} {Phys.\ Rev.\ D}\ }\textbf {\bibinfo {volume} {75}},\ \bibinfo
  {pages} {123509} (\bibinfo {year} {2007}{\natexlab{b}})},\ \Eprint
  {http://arxiv.org/abs/astro-ph/0702002} {arXiv:astro-ph/0702002} \BibitemShut
  {NoStop}%
\bibitem [{\citenamefont {Lake}\ and\ \citenamefont
  {Abdelqader}(2011)}]{Lake:2011ni}%
  \BibitemOpen
  \bibfield  {author} {\bibinfo {author} {\bibfnamefont {K.}~\bibnamefont
  {Lake}}\ and\ \bibinfo {author} {\bibfnamefont {M.}~\bibnamefont
  {Abdelqader}},\ }\href {\doibase 10.1103/PhysRevD.84.044045} {\bibfield
  {journal} {\bibinfo  {journal} {Phys.\ Rev.\ D}\ }\textbf {\bibinfo {volume}
  {84}},\ \bibinfo {pages} {044045} (\bibinfo {year} {2011})},\ \Eprint
  {http://arxiv.org/abs/1106.3666} {arXiv:1106.3666 [gr-qc]} \BibitemShut
  {NoStop}%
\bibitem [{\citenamefont {Faraoni}\ and\ \citenamefont
  {Jacques}(2007)}]{Faraoni:2007es}%
  \BibitemOpen
  \bibfield  {author} {\bibinfo {author} {\bibfnamefont {V.}~\bibnamefont
  {Faraoni}}\ and\ \bibinfo {author} {\bibfnamefont {A.}~\bibnamefont
  {Jacques}},\ }\href {\doibase 10.1103/PhysRevD.76.063510} {\bibfield
  {journal} {\bibinfo  {journal} {Phys.\ Rev.\ D}\ }\textbf {\bibinfo {volume}
  {76}},\ \bibinfo {pages} {063510} (\bibinfo {year} {2007})},\ \Eprint
  {http://arxiv.org/abs/0707.1350} {arXiv:0707.1350 [gr-qc]} \BibitemShut
  {NoStop}%
\bibitem [{\citenamefont {Carrera}\ and\ \citenamefont
  {Giulini}(2010)}]{Carrera:2009ve}%
  \BibitemOpen
  \bibfield  {author} {\bibinfo {author} {\bibfnamefont {M.}~\bibnamefont
  {Carrera}}\ and\ \bibinfo {author} {\bibfnamefont {D.}~\bibnamefont
  {Giulini}},\ }\href {\doibase 10.1103/PhysRevD.81.043521} {\bibfield
  {journal} {\bibinfo  {journal} {Phys.\ Rev.\ D}\ }\textbf {\bibinfo {volume}
  {81}},\ \bibinfo {pages} {043521} (\bibinfo {year} {2010})},\ \Eprint
  {http://arxiv.org/abs/0908.3101} {arXiv:0908.3101 [gr-qc]} \BibitemShut
  {NoStop}%
\bibitem [{\citenamefont {Faraoni}\ \emph {et~al.}(2012)\citenamefont
  {Faraoni}, \citenamefont {{Zambrano Moreno}},\ and\ \citenamefont
  {Nandra}}]{Faraoni:2012gz}%
  \BibitemOpen
  \bibfield  {author} {\bibinfo {author} {\bibfnamefont {V.}~\bibnamefont
  {Faraoni}}, \bibinfo {author} {\bibfnamefont {A.~F.}\ \bibnamefont {{Zambrano
  Moreno}}}, \ and\ \bibinfo {author} {\bibfnamefont {R.}~\bibnamefont
  {Nandra}},\ }\href {\doibase 10.1103/PhysRevD.85.083526} {\bibfield
  {journal} {\bibinfo  {journal} {Phys.\ Rev.\ D}\ }\textbf {\bibinfo {volume}
  {85}},\ \bibinfo {pages} {083526} (\bibinfo {year} {2012})},\ \Eprint
  {http://arxiv.org/abs/1202.0719} {arXiv:1202.0719 [gr-qc]} \BibitemShut
  {NoStop}%
\bibitem [{\citenamefont {da~Silva}\ \emph {et~al.}(2013)\citenamefont
  {da~Silva}, \citenamefont {Fontanini},\ and\ \citenamefont
  {Guariento}}]{daSilva:2012nh}%
  \BibitemOpen
  \bibfield  {author} {\bibinfo {author} {\bibfnamefont {A.~M.}\ \bibnamefont
  {da~Silva}}, \bibinfo {author} {\bibfnamefont {M.}~\bibnamefont {Fontanini}},
  \ and\ \bibinfo {author} {\bibfnamefont {D.~C.}\ \bibnamefont {Guariento}},\
  }\href {\doibase 10.1103/PhysRevD.87.064030} {\bibfield  {journal} {\bibinfo
  {journal} {Phys.\ Rev.\ D}\ }\textbf {\bibinfo {volume} {87}},\ \bibinfo
  {pages} {064030} (\bibinfo {year} {2013})},\ \Eprint
  {http://arxiv.org/abs/1212.0155} {arXiv:1212.0155 [gr-qc]} \BibitemShut
  {NoStop}%
\bibitem [{\citenamefont {Raychaudhuri}(1979)}]{raychaudhuri}%
  \BibitemOpen
  \bibfield  {author} {\bibinfo {author} {\bibfnamefont {A.~K.}\ \bibnamefont
  {Raychaudhuri}},\ }\href@noop {} {\emph {\bibinfo {title} {Theoretical
  Cosmology}}},\ Oxford Studies in Physics\ (\bibinfo  {publisher} {Clarendon
  Press},\ \bibinfo {address} {Oxford},\ \bibinfo {year} {1979})\BibitemShut
  {NoStop}%
\bibitem [{\citenamefont {Mart\'{\i}nez}\ \emph {et~al.}(2006)\citenamefont
  {Mart\'{\i}nez}, \citenamefont {Troncoso},\ and\ \citenamefont
  {Staforelli}}]{Martinez:2005di}%
  \BibitemOpen
  \bibfield  {author} {\bibinfo {author} {\bibfnamefont {C.}~\bibnamefont
  {Mart\'{\i}nez}}, \bibinfo {author} {\bibfnamefont {R.}~\bibnamefont
  {Troncoso}}, \ and\ \bibinfo {author} {\bibfnamefont {J.~P.}\ \bibnamefont
  {Staforelli}},\ }\href {\doibase 10.1103/PhysRevD.74.044028} {\bibfield
  {journal} {\bibinfo  {journal} {Phys.\ Rev.\ D}\ }\textbf {\bibinfo {volume}
  {74}},\ \bibinfo {pages} {044028} (\bibinfo {year} {2006})},\ \Eprint
  {http://arxiv.org/abs/hep-th/0512022} {arXiv:hep-th/0512022} \BibitemShut
  {NoStop}%
\bibitem [{\citenamefont {Brill}\ and\ \citenamefont
  {Hayward}(1994)}]{Brill:1993tw}%
  \BibitemOpen
  \bibfield  {author} {\bibinfo {author} {\bibfnamefont {D.~R.}\ \bibnamefont
  {Brill}}\ and\ \bibinfo {author} {\bibfnamefont {S.~A.}\ \bibnamefont
  {Hayward}},\ }\href {\doibase 10.1088/0264-9381/11/2/008} {\bibfield
  {journal} {\bibinfo  {journal} {Classical Quantum Gravity}\ }\textbf
  {\bibinfo {volume} {11}},\ \bibinfo {pages} {359} (\bibinfo {year} {1994})},\
  \Eprint {http://arxiv.org/abs/gr-qc/9304007} {arXiv:gr-qc/9304007}
  \BibitemShut {NoStop}%
\bibitem [{\citenamefont {Babichev}\ \emph {et~al.}(2006)\citenamefont
  {Babichev}, \citenamefont {Mukhanov},\ and\ \citenamefont
  {Vikman}}]{Babichev:2006vx}%
  \BibitemOpen
  \bibfield  {author} {\bibinfo {author} {\bibfnamefont {E.}~\bibnamefont
  {Babichev}}, \bibinfo {author} {\bibfnamefont {V.}~\bibnamefont {Mukhanov}},
  \ and\ \bibinfo {author} {\bibfnamefont {A.}~\bibnamefont {Vikman}},\ }\href
  {\doibase 10.1088/1126-6708/2006/09/061} {\bibfield  {journal} {\bibinfo
  {journal} {J.\ High Energy Phys.\@}\ }\textbf {\bibinfo {volume} {2006}},\
  \bibinfo {pages} {061} (\bibinfo {year} {2006})},\ \Eprint
  {http://arxiv.org/abs/hep-th/0604075} {arXiv:hep-th/0604075} \BibitemShut
  {NoStop}%
\bibitem [{\citenamefont {Babichev}\ \emph {et~al.}(2008)\citenamefont
  {Babichev}, \citenamefont {Mukhanov},\ and\ \citenamefont
  {Vikman}}]{Babichev:2007dw}%
  \BibitemOpen
  \bibfield  {author} {\bibinfo {author} {\bibfnamefont {E.}~\bibnamefont
  {Babichev}}, \bibinfo {author} {\bibfnamefont {V.}~\bibnamefont {Mukhanov}},
  \ and\ \bibinfo {author} {\bibfnamefont {A.}~\bibnamefont {Vikman}},\ }\href
  {\doibase 10.1088/1126-6708/2008/02/101} {\bibfield  {journal} {\bibinfo
  {journal} {J.\ High Energy Phys.\@}\ }\textbf {\bibinfo {volume} {2008}},\
  \bibinfo {pages} {101} (\bibinfo {year} {2008})},\ \Eprint
  {http://arxiv.org/abs/0708.0561} {arXiv:0708.0561 [hep-th]} \BibitemShut
  {NoStop}%
\bibitem [{\citenamefont {Vikman}(2007)}]{vikman-2007}%
  \BibitemOpen
  \bibfield  {author} {\bibinfo {author} {\bibfnamefont {A.}~\bibnamefont
  {Vikman}},\ }\emph {\bibinfo {title} {K-essence: cosmology, causality and
  emergent geometry}},\ \href {http://edoc.ub.uni-muenchen.de/7761/} {Ph.D.
  thesis},\ \bibinfo  {school} {Ludwig-Maximilians-Universit\"at}, \bibinfo
  {address} {M\"unchen} (\bibinfo {year} {2007})\BibitemShut {NoStop}%
\bibitem [{\citenamefont {Misra}\ and\ \citenamefont
  {Srivastava}(1973)}]{Misra:1973zz}%
  \BibitemOpen
  \bibfield  {author} {\bibinfo {author} {\bibfnamefont {R.~M.}\ \bibnamefont
  {Misra}}\ and\ \bibinfo {author} {\bibfnamefont {D.~C.}\ \bibnamefont
  {Srivastava}},\ }\href {\doibase 10.1103/PhysRevD.8.1653} {\bibfield
  {journal} {\bibinfo  {journal} {Phys.\ Rev.\ D}\ }\textbf {\bibinfo {volume}
  {8}},\ \bibinfo {pages} {1653} (\bibinfo {year} {1973})}\BibitemShut
  {NoStop}%
\bibitem [{\citenamefont {Ho\v{r}ava}(2009)}]{horava-prd-2009}%
  \BibitemOpen
  \bibfield  {author} {\bibinfo {author} {\bibfnamefont {P.}~\bibnamefont
  {Ho\v{r}ava}},\ }\href {\doibase 10.1103/PhysRevD.79.084008} {\bibfield
  {journal} {\bibinfo  {journal} {Phys.\ Rev.\ D}\ }\textbf {\bibinfo {volume}
  {79}},\ \bibinfo {pages} {084008} (\bibinfo {year} {2009})},\ \Eprint
  {http://arxiv.org/abs/0901.3775} {arXiv:0901.3775 [hep-th]} \BibitemShut
  {NoStop}%
\bibitem [{\citenamefont {Afshordi}(2009)}]{Afshordi:2009tt}%
  \BibitemOpen
  \bibfield  {author} {\bibinfo {author} {\bibfnamefont {N.}~\bibnamefont
  {Afshordi}},\ }\href {\doibase 10.1103/PhysRevD.80.081502} {\bibfield
  {journal} {\bibinfo  {journal} {Phys.\ Rev.\ D}\ }\textbf {\bibinfo {volume}
  {80}},\ \bibinfo {pages} {081502} (\bibinfo {year} {2009})},\ \Eprint
  {http://arxiv.org/abs/0907.5201} {arXiv:0907.5201 [hep-th]} \BibitemShut
  {NoStop}%
\bibitem [{\citenamefont {Gomes}\ \emph {et~al.}(2011)\citenamefont {Gomes},
  \citenamefont {Gryb},\ and\ \citenamefont {Koslowski}}]{Gomes:2010fh}%
  \BibitemOpen
  \bibfield  {author} {\bibinfo {author} {\bibfnamefont {H.}~\bibnamefont
  {Gomes}}, \bibinfo {author} {\bibfnamefont {S.}~\bibnamefont {Gryb}}, \ and\
  \bibinfo {author} {\bibfnamefont {T.}~\bibnamefont {Koslowski}},\ }\href
  {\doibase 10.1088/0264-9381/28/4/045005} {\bibfield  {journal} {\bibinfo
  {journal} {Classical Quantum Gravity}\ }\textbf {\bibinfo {volume} {28}},\
  \bibinfo {pages} {045005} (\bibinfo {year} {2011})},\ \Eprint
  {http://arxiv.org/abs/1010.2481} {arXiv:1010.2481 [gr-qc]} \BibitemShut
  {NoStop}%
\end{thebibliography}%

\end{document}